\documentclass[conference]{IEEEtran}
\IEEEoverridecommandlockouts

\usepackage{cite}
\usepackage{amsmath,amssymb,amsfonts}
\usepackage{algorithmic}
\usepackage{graphicx}
\usepackage{textcomp}
\usepackage{xcolor}
\usepackage{hyperref}
\usepackage{subcaption}

\newcommand{\transpose}[1]{\ensuremath{{#1}^{\textsc{t}}}}

\def\BibTeX{{\rm B\kern-.05em{\sc i\kern-.025em b}\kern-.08em
    T\kern-.1667em\lower.7ex\hbox{E}\kern-.125emX}}
\begin{document}

\title{Sample-level EEG-based Selective Auditory Attention Decoding with Markov Switching Models\\

\thanks{This research is funded by the Research Foundation - Flanders (FWO) project No G081722N, junior postdoctoral fellowship fundamental research of the FWO (for S. Geirnaert, No. 1242524N),  Internal Funds KU Leuven (projects IDN/23/006, C14/25/108, and C3/25/107), and the Flemish Government (AI Research Program). \\All authors are also affiliated with Leuven.AI - KU Leuven institute for AI, Belgium.}
}

\author{\IEEEauthorblockN{Yuanyuan Yao}
\IEEEauthorblockA{\textit{Dept. of Electrical Engineering, STADIUS} \\
\textit{KU Leuven}\\
Leuven, Belgium \\
yuanyuan.yao@kuleuven.be}
\and
\IEEEauthorblockN{Simon Geirnaert}
\IEEEauthorblockA{\textit{Dept. of Electrical Engineering, STADIUS} \\
\textit{Dept. of Neurosciences, ExpORL} \\
\textit{KU Leuven}\\
Leuven, Belgium \\
simon.geirnaert@kuleuven.be}
\and[\hfill\mbox{}\par\mbox{}\hfill]
\IEEEauthorblockN{Tinne Tuytelaars}
\IEEEauthorblockA{\textit{Dept. of Electrical Engineering, PSI} \\
\textit{KU Leuven}\\
Leuven, Belgium \\
Tinne.Tuytelaars@kuleuven.be}
\and
\IEEEauthorblockN{Alexander Bertrand}
\IEEEauthorblockA{\textit{Dept. of Electrical Engineering, STADIUS} \\
\textit{KU Leuven}\\
Leuven, Belgium \\
Alexander.Bertrand@kuleuven.be}
}

\maketitle

\begin{abstract}
Selective auditory attention decoding aims to identify the speaker of interest from listeners' neural signals, such as electroencephalography (EEG), in the presence of multiple concurrent speakers. Most existing methods operate at the window level, facing a trade-off between temporal resolution and decoding accuracy. Recent work has shown that hidden Markov model (HMM)-based post-processing can smooth window-level decoder outputs to improve this trade-off. Instead of using a separate smoothing step, we propose to integrate the decoding and smoothing components into a single probabilistic framework using a Markov switching model (MSM). It directly models the relationship between the EEG and speech envelopes under each attention state while incorporating the temporal dynamics of attention. This formulation enables sample-level attention decoding, with model parameters and attention states jointly estimated via the expectation-maximization algorithm. Experimental results demonstrate that this integrated MSM formulation achieves comparable decoding accuracy to HMM post-processing while providing faster attention switch detection. The code for the proposed method is available at \url{https://github.com/YYao-42/MSM}.
\end{abstract}

\begin{IEEEkeywords}
auditory attention decoding, electroencephalography, Markov Switching Model
\end{IEEEkeywords}

\section{Introduction}

\par Real-life social situations often involve multiple concurrent speakers, where listeners use selective attention to focus on a single speaker of interest and suppress interfering sound sources \cite{cherry1953some}. Tracking the attended speaker from neural signals, known as selective auditory attention decoding (sAAD), has been an active research topic in recent years, with potential applications in neuro-steered hearing aids \cite{geirnaert2021eegBased}. 

\par Electroencephalography (EEG) is a popular choice for recording neural activity in sAAD due to its non-invasiveness and high temporal resolution. However, non-invasiveness comes at the cost of an extremely low signal-to-noise ratio (SNR), making it necessary to aggregate information over time for more reliable decoding. Therefore, the vast majority of sAAD algorithms decode attention based on aggregated EEG data within a pre-defined window length, thereby not fully exploiting the high temporal resolution offered by EEG. A common framework involves reconstructing the attended speech envelope from neural signals using a linear decoder and then comparing it to the speech envelopes of the candidate speakers \cite{geirnaert2021eegBased, o2015attentional, biesmans2016auditory, alickovic2019tutorial}. By calculating the correlation between the reconstructed signal and each candidate speaker envelope over a fixed time window, the speaker with the highest correlation is identified as the attended one. More recently, deep learning models have streamlined this by directly mapping windows of neural and speech data to attention metrics such as correlations and probabilities \cite{ciccarelli2019comparison, puffay2023relating, nguyen2024aadnet}. However, both approaches face a fundamental trade-off: while shorter windows offer higher temporal resolution, they also produce noisier correlations or probabilities, leading to degraded decoding performance. This motivates methods that can smooth these outputs over time to achieve higher decoding accuracy with minimal impact on the temporal resolution.

\par To address this, Heintz et al. \cite{heintz2025post} proposed a post-processing method to refine raw sAAD output metrics, such as correlations, using a hidden Markov model (HMM), which models these outputs as noisy observations conditioned on an underlying attention state. The idea is to exploit the fact that human attention does not switch very frequently, modeling the attention state as a Markov chain where the current focus is likely to persist in the next moment. By assigning a high self-transition probability to the current state, the method can effectively smooth out spurious fluctuations in the raw sAAD outputs on 1-second windows and achieve comparable decoding performance to that of 30-second windows at a significantly better temporal resolution. 

\par Inspired by this work, we propose to model sAAD as a Markov switching model (MSM), which is an integrated framework that not only incorporates the temporal structure of attention as in HMMs but also directly models the relationship between EEG and speech envelope under each attention state. The expectation-maximization (EM) algorithm is used to jointly estimate the model parameters and decode the attention state at each time point. Therefore, instead of operating as a post-processing step at the window level, the MSM can be directly applied to the EEG and speech data to decode attention at the sample level. We find that the MSM can achieve comparable decoding accuracy to the HMM post-processing method based on correlations generated by a linear decoder, while achieving even shorter switch detection delay.

\section{Markov Switching Model}

\par The Markov switching model (MSM) was first proposed by Hamilton \cite{hamilton1990analysis} to model economic time series with regime changes. It assumes each observation $y_t$ is drawn from one of several regimes, each characterized by its own set of parameters. The regime at time $t$ is determined by a hidden state variable $S_t$ that evolves according to a Markov chain. In the context of sAAD, considering a two-speaker setting, $S_t$ indicates which speaker is attended at time $t$, with $S_t=1$ for speaker 1 and $S_t=2$ for speaker 2. Denote the speech envelopes of speaker 1 and speaker 2 at time $t$ as $y_{1t}$ and $y_{2t}$, respectively. The goal of sAAD is to decode $S_t$ from the recorded EEG signals $\mathbf x_t$ at each time point $t$. We augment each EEG sample $\mathbf x_t$ with $L-1$ time lags to form a $CL$-dimensional spatial-temporal vector $\hat{\mathbf x}_t$, where $C$ is the number of EEG channels:
\begin{equation}
  \begin{aligned}
    \hat {\mathbf x}_t = [&x_{1,t}, x_{1,t-1}, \ldots, x_{1,t-L+1}, \ldots,\\&  x_{C,t}, x_{C,t-1}, \ldots, x_{C,t-L+1} ]^{\textsc{t}}.
  \end{aligned}
\end{equation}

\par We define the observation $y_t= y_{1t} - y_{2t}$ to reduce the problem to a one-dimensional observation. We assume the following model for $y_t$, conditioned on the hidden state variable $S_t$:
\begin{equation}
y_t = y_{1t}-y_{2t} = \transpose{\boldsymbol \beta}_{S_t} \hat{\mathbf x}_t + e_t,\quad e_t \sim \mathcal N(0,\sigma_{S_t}^2),
\label{eq:def_MSM}
\end{equation}
where ${\boldsymbol \beta}_{S_t}$ and $\sigma_{S_t}^2$ are the regression coefficients and noise variance corresponding to the attention state $S_t$. Here, ${\boldsymbol \beta}_{S_t}$ is akin to the traditional linear decoder often used in sAAD \cite{geirnaert2021eegBased, o2015attentional, biesmans2016auditory}, yet now it decodes the difference between speech envelopes rather than each speech envelope separately\footnote{One could also use a bivariate model for $\transpose{[y_{1t}~ y_{2t}]}$ by making $\boldsymbol{\beta}_{S_t}$ a $2 \times CL$ matrix. The univariate model \eqref{eq:def_MSM} can then be viewed as a subtraction of the two rows of this bivariate system. This univariate formulation was observed to behave more stably during training.}.

\par The attention state $S_t$ evolves as a first-order Markov chain with transition probabilities
\begin{equation}
 p_{ij} = \Pr(S_t=j| S_{t-1}=i), \quad i,j\in\{1,2\},  
 \label{eq:trans_prob}
\end{equation}
with $\sum_{j\in\{1,2\}} p_{ij} = 1$ for each $i$. These transition probabilities describe how frequently attention switches between speakers, as in the HMM \cite{heintz2025post}. Similar to the HMM approach, we set $p_{11}=p_{22}$ as a tunable parameter for controlling the temporal smoothness of the decoded attention state sequence to reduce the number of hyperparameters and ensure symmetric treatment of both speakers.

\par The attention state sequence and the unknown parameters are to be jointly estimated such that the likelihood of the observed data is maximized. Since the attention state sequence is hidden, we use the expectation-maximization (EM) algorithm to iteratively estimate the parameters and decode the attention states \cite{dempster1977maximum}, as explained below.

\subsection{Forward-backward algorithm}

\par Let $\boldsymbol \psi_t = \{y_1, ..., y_t\}$ denote the observed data up to time $t$, and $\boldsymbol{\Theta} = \{\boldsymbol \beta_1, \boldsymbol \beta_2, \sigma_1^2, \sigma_2^2\}$ denote the set of to-be-estimated model parameters. In the EM algorithm, we need to compute the posterior probabilities of the hidden states given the observed data and current parameter estimates, i.e., $p(S_t|\boldsymbol \psi_T;\boldsymbol \Theta^k), t=1,... T$, where $\boldsymbol \Theta^k$ denotes the parameter estimates at the $k$-th iteration and is omitted for brevity in the following equations. This can be computed using the forward-backward algorithm \cite{kim2017state}, with
\begin{itemize}
    \item \textit{Forward pass:} Computing the posterior probabilities $p(S_t|\boldsymbol \psi_t)$ with two iterative steps:
    \begin{align}        
        \operatorname{Pr}(S_{t}| \boldsymbol\psi_{t})&= \operatorname{Pr}(S_{t}| \boldsymbol\psi_{t-1},y_t)=\frac{f(S_{t},y_t| \boldsymbol\psi_{t-1})}{f(y_t| \boldsymbol\psi_{t-1})}\nonumber\\&= \frac{f(y_t| S_{t} )\operatorname{Pr}(S_{t}|  \boldsymbol\psi_{t-1})}{\sum_{S_t}f(y_t| S_{t})\operatorname{Pr}(S_{t}|  \boldsymbol\psi_{t-1})}, \label{eq:fw_1}\\
        \operatorname{Pr}(S_t| \boldsymbol\psi_{t-1})&= \sum_{S_{t-1}}\operatorname{Pr}(S_t| S_{t-1})\operatorname{Pr}(S_{t-1}| \boldsymbol\psi_{t-1}),\label{eq:fw_2}
    \end{align}
    where $f(y_t| S_t)$ is the likelihood of observation $y_t$ given state $S_t$ that can be computed from \eqref{eq:def_MSM}, and $\operatorname{Pr}(S_t| S_{t-1})$ is given by \eqref{eq:trans_prob}. $\operatorname{Pr}(S_0|\boldsymbol\psi_{0})=\operatorname{Pr}(S_0)$ is the initial state distribution and can be set to a uniform distribution if no prior knowledge is available.
    \item \textit{Backward pass:} Given $\operatorname{Pr}(S_T|\boldsymbol\psi_{T})$ at the last iteration of the forward pass, we can compute the smoothing probabilities $\operatorname{Pr}( S_{t}|\boldsymbol \psi_{T})$ iteratively backward in time:
    \begin{align}&\operatorname{Pr}( S_{t}|\boldsymbol \psi_{T})=\sum_{S_{t+1}}\operatorname{Pr}( S_{t}, S_{t+1} |\boldsymbol \psi_{T}) \nonumber\\ =&\sum_{S_{t+1}}\operatorname{Pr}( S_{t} | S_{t+1}, \boldsymbol \psi_{t})\operatorname{Pr}( S_{t+1} | \boldsymbol \psi_{T}) \nonumber \\=& \sum_{S_{t+1}} \frac{\operatorname{Pr}( S_{t},  S_{t+1}| \boldsymbol \psi_{t})\operatorname{Pr}( S_{t+1} | \boldsymbol \psi_{T})}{\operatorname{Pr}(S_{t+1}| \boldsymbol \psi_{t})} \nonumber \\=& \sum_{S_{t+1}} \frac{\operatorname{Pr}( S_{t},  S_{t+1}| \boldsymbol \psi_{t})\operatorname{Pr}( S_{t+1} | \boldsymbol \psi_{T})}{\sum_{S_t}\operatorname{Pr}( S_{t},  S_{t+1}| \boldsymbol \psi_{t})}\nonumber \\=& \sum_{S_{t+1}} \frac{\operatorname{Pr}( S_{t} | \boldsymbol \psi_{t})\operatorname{Pr}( S_{t+1} | S_t)\operatorname{Pr}( S_{t+1} | \boldsymbol \psi_{T})}{\sum_{S_t}[\operatorname{Pr}( S_{t} | \boldsymbol \psi_{t})\operatorname{Pr}( S_{t+1} | S_t)]}, \end{align}
    where $\operatorname{Pr}( S_{t} | \boldsymbol \psi_{t})$ is provided by the forward pass \eqref{eq:fw_1}-\eqref{eq:fw_2} and $\operatorname{Pr}( S_{t+1} | S_t)$ by \eqref{eq:trans_prob}.
\end{itemize}
Note that here we assume the full observed data are available to improve the inference of $S_t$ using future information. However, the method can easily be made causal by performing only the forward pass.

\subsection{EM iterations}

\par The EM algorithm performs two steps iteratively until convergence. In the \textit{E-step}, we formulate the Q-function, which is the expectation of the complete data log-likelihood w.r.t. the posterior distribution of the hidden states given the observed data and current parameter estimates:
\begin{equation}
    Q_k(\boldsymbol \Theta)=E_{\boldsymbol{\xi}_T|\boldsymbol \psi_T;\boldsymbol \Theta^k}[\log f(\boldsymbol{\xi}_T,\boldsymbol \psi_T;\boldsymbol \Theta)],
\end{equation}
where $\boldsymbol{\xi}_T = \{S_1,...,S_T\}$ is the sequence of hidden states. The likelihood of the complete data can be factorized as

\begin{align}
    \label{eq:factorization}
    f(\boldsymbol{\xi}_T,\boldsymbol \psi_T;\boldsymbol \Theta)&= f(\boldsymbol \psi_T| \boldsymbol{\xi}_T;\boldsymbol \Theta)\operatorname{Pr}(\boldsymbol{\xi}_T)\nonumber \\
    &=\prod_{t} f(y_t| S_t;\boldsymbol \Theta)\prod_{t} \operatorname{Pr}(S_t| S_{t-1}), 
\end{align}
and thus the Q-function can be written as
\begin{equation}
    Q_k(\boldsymbol \Theta)=\sum_{t} \sum_{S_t} \operatorname{Pr}(S_t|\boldsymbol \psi_T;\boldsymbol \Theta^k) \log f(y_t| S_t;\boldsymbol \Theta) + \mathcal C,
\end{equation}
where $\mathcal C$ is a constant independent of $\boldsymbol \Theta$ coming from the factor $\prod_{t} \operatorname{Pr}(S_t| S_{t-1})$ in \eqref{eq:factorization}.

\par In the \textit{M-step}, we set the derivatives of the Q-function w.r.t. the model parameters to zero and solve for the parameters that maximize the Q-function. It is straightforward to derive the update rules for the regression coefficients and noise variances \cite{kim2017state}:
\begin{align} 
    &\boldsymbol{\beta}_{i}^{k+1} = \nonumber \\& [\sum_t \hat{\mathbf x}_t \transpose{\hat{\mathbf x}_t} \operatorname{Pr}(S_t=i|\boldsymbol \psi_T;\boldsymbol \Theta^k)]^{-1} [\sum_t \hat{\mathbf x}_t y_t \operatorname{Pr}(S_t=i|\boldsymbol \psi_T;\boldsymbol \Theta^k)],\\
    &(\sigma_{i}^{k+1})^{2} = \frac{\sum_t (y_t - \transpose{\hat{\mathbf x}_t} \boldsymbol{\beta}_{i}^{k+1})^2 \operatorname{Pr}(S_t=i|\boldsymbol \psi_T;\boldsymbol \Theta^k)}{\sum_t \operatorname{Pr}(S_t=i|\boldsymbol \psi_T;\boldsymbol \Theta^k)}
\end{align}
for $i \in \{1,2\}$. The inference of the hidden states $\operatorname{Pr}(S_t|\boldsymbol \psi_T;\boldsymbol \Theta^k)$ in each iteration is performed using the forward-backward algorithm described in the previous subsection.

\subsection{Initialization}
\label{sec:init}

\par Initialization is important for the convergence of the EM algorithm, especially when working with low SNR data such as EEG. Therefore, although the proposed method can in principle be initialized with random parameters and work in a purely unsupervised manner, we find that it greatly accelerates the convergence and leads to better decoding performance if we initialize the parameters with a pretrained decoder $\boldsymbol{\beta}^*$ that is trained, e.g., on a separate dataset using linear regression as in \cite{geirnaert2021eegBased, o2015attentional, biesmans2016auditory}, where the attended speaker is used as the target. Intuitively, in our formulation \eqref{eq:def_MSM}, the state-dependent decoders $\boldsymbol{\beta}_1$ and $\boldsymbol{\beta}_2$ can be initialized as $\boldsymbol{\beta}^*$ and $-\boldsymbol{\beta}^*$, respectively, assuming $\hat{\mathbf x}_t$ is mainly dominated by the neural response to the attended speaker\footnote{One could also subtract a pre-trained unattended-envelope decoder from $\boldsymbol{\beta}^*$ to make the initialization more compatible with \eqref{eq:def_MSM}, but this did not improve the results.}. The noise variances $\sigma_1^2$ and $\sigma_2^2$ can have the same initialization, e.g., the mean squared error of the pretrained decoder $\boldsymbol{\beta}^*$ on the training data, if we assume the speech envelopes are properly normalized and the reconstruction error is similar regardless of which speaker is attended. We note that this symmetry is only used for initialization, but it is not enforced within the EM algorithm during training, as the underlying assumptions could be violated in practice.

\begin{figure*}[htbp]
    \centering
    \begin{subfigure}[b]{0.7\textwidth}
        \centering
        \includegraphics[width=\textwidth]{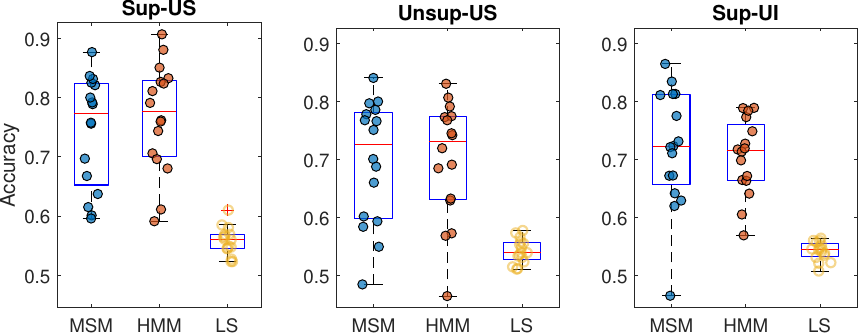}
        \caption{Decoding accuracy.}
        \label{fig:acc}
    \end{subfigure}
    \hfill 
    \begin{subfigure}[b]{0.7\textwidth}
        \centering
        \includegraphics[width=\textwidth]{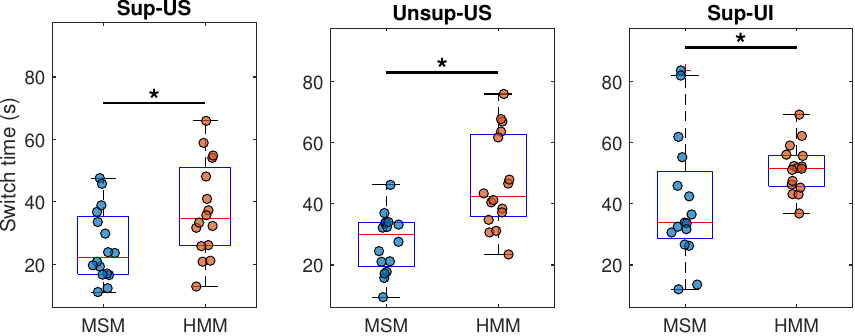}
        \caption{Switch detection time.}
        \label{fig:switch_time}
    \end{subfigure}
    \caption{Performance comparison of the proposed MSM method and the HMM post-processing method (operating on 1-second windows) under different settings. The dots mark the cross-validated outcomes for each participant. Box-and-whisker plots are used to display the median and quartiles, with the whiskers encompassing the full data spread excluding outliers. ``*" indicates the results of MSM are significantly lower than HMM (Wilcoxon signed-rank test, $p<0.05$). The decoding accuracy of the LS decoder without post-processing is also shown for reference. Its switch detection time is not reported, as the noisy outputs produce a highly fragmented state sequence, which does not allow computing a meaningful switch detection time.}
    \label{fig:combined_performance}
\end{figure*}

\section{Experiment}

\par We compare the proposed MSM method with the HMM post-processing method (non-causal version) applied to neural tracking correlations generated by a linear decoder as proposed by Heintz et al. \cite{heintz2025post} on a dataset \cite{biesmans2016auditory} that is publicly available at \cite{das_2019_4004271}. The dataset contains 72-min audio data with two competing speakers and the corresponding 64-channel EEG data from 16 normal-hearing participants instructed to attend to one of the speakers in each trial. The speech envelopes are extracted using a gammatone filterbank as described in Biesmans et al. \cite{biesmans2016auditory}. The audio and EEG data are bandpass filtered in the 0.1-4 Hz range \cite{li2025temporal} and downsampled to 10 Hz. To artificially increase the number of speaker switches, we divide the data into 1-minute segments and randomly swap the assignment of speaker 1 and speaker 2 in each segment. For MSM, the performance is evaluated directly on the per-sample decoding results without any window-based post-processing.

\subsection{Metrics and parameters}

\par We define the decoding accuracy as the percentage of time points where the predicted attended speaker matches the true attended speaker. The switch detection time is defined as the time difference between the true attention switch and the nearest time point where the decoded attention state switches correctly, i.e., the posterior probability for the newly attended speaker exceeds $50\%$. Since both methods are non-causal, meaning that this can happen before the actual switch, we take the absolute value of the switch detection time and average it across all true attention switches. For the HMM post-processing method, switch detection time is by definition larger than the chosen window length as window-based processing has this inherent algorithmic delay. If the computed switch detection time is larger than the time interval between two true attention switches, this is considered a missed switch and the switch detection time is set to the time interval between two true attention switches.

\par The comparison is performed under three settings based on how the pretrained decoder $\boldsymbol \beta^*$ is obtained:
\begin{itemize}
    \item \textit{Supervised user-specific (Sup-US):} $\boldsymbol \beta^*$ is trained for each participant with attention labels (on a held-out part of the data of the participant) using least squares (LS), as commonly done in the literature \cite{geirnaert2021eegBased, o2015attentional, biesmans2016auditory}.
    \item \textit{Unsupervised user-specific (Unsup-US):} $\boldsymbol \beta^*$ is trained for each participant in an unsupervised manner as proposed by Geirnaert et al. in \cite{geirnaert2021unsupervised}.
    \item \textit{Supervised user-independent (Sup-UI):} $\boldsymbol \beta^*$ is trained as in the \textit{Sup-US} setting but with data from all participants (except the test participant) concatenated together.
\end{itemize}
For all settings, $\boldsymbol \beta^*$ is a spatial-temporal decoder that captures EEG information from $0-500$ ms after stimulus onset. The obtained pretrained decoder $\boldsymbol \beta^*$ is used to initialize the MSM as described in Section \ref{sec:init}, and the HMM post-processing method is applied to the outputs of the pretrained decoder. In the \textit{Sup-US} and \textit{Unsup-US} settings, the pretrained decoder is cross-validated on the same participant's data (3 folds), while in the \textit{Sup-UI} setting, we use leave-one-participant-out cross-validation. In both cases, $\boldsymbol\beta^*$ is trained on the available training data (either 2 folds or $N-1$ participants). While the test set is used solely for evaluation in the HMM method, the MSM method utilizes the test data for two purposes: first, to fit the MSM parameters via the EM algorithm, and subsequently to evaluate decoding performance. Note that no attention labels are needed during EM fitting, so this does not introduce label leakage.

\par As pointed out by Heintz et al. \cite{heintz2025post}, the transition probabilities $p_{ij}$ and the window length are important parameters that influence the decoding accuracy and switch detection time. Therefore, we follow their setting and evaluate on correlations obtained from 1-second windows with $p_{12}=p_{21}=10^{-3}$. For a fair comparison, we set the MSM's transition probabilities as $p_{12}=p_{21}=10^{-4}$ such that the expected number of switches per unit time is equal for both methods, given the $10$ Hz sampling frequency. In the original HMM post-processing method, the distributions of the correlation between the reconstructed speech envelope and the ground-truth speech envelope under attended and unattended states are estimated in a supervised manner. These distributions are then used when computing the emission probabilities. To ensure a fair comparison with the MSM, which does not require attention labels, we modify the HMM's emission distribution estimation to be unsupervised: we pool the correlations from both states and fit a two-component Gaussian mixture model where the component with the higher mean is assigned to the attended state.

\subsection{Results}

\par We present the decoding accuracy and switch detection time for the proposed MSM and the HMM post-processing methods across the three initialization settings (\textit{Sup-US}, \textit{Unsup-US}, and \textit{Sup-UI}) in Fig. \ref{fig:combined_performance}. For reference, the decoding accuracy using raw correlations on 1-second windows without HMM post-processing (denoted as ``LS") is included in Fig. \ref{fig:acc}. Under all settings, the accuracy of the raw LS decoder ranges between $50\%$ and $60\%$, which is near chance-level. With HMM post-processing, the median decoding accuracy improves to $77.6\%$, $73.1\%$, and $71.6\%$ for the \textit{Sup-US}, \textit{Unsup-US}, and \textit{Sup-UI} settings, respectively. The MSM achieves comparable results, with median accuracies of $77.3\%$, $72.6\%$, and $72.2\%$, respectively. No significant difference in decoding accuracy was observed between the two methods across all settings (Wilcoxon signed-rank test, $p > 0.05$ in all cases).

\par Regarding switch detection time (Fig. \ref{fig:switch_time}), the LS decoder is excluded because its noisy outputs lead to highly fragmented state sequences, making the switch detection time poorly defined. The MSM achieves significantly shorter switch detection times than the HMM post-processing method across all settings ($p = 0.0004$, $0.0003$, and $0.0233$ for \textit{Sup-US}, \textit{Unsup-US}, and \textit{Sup-UI}, respectively). Specifically, the median switch detection time drops from $34.5$ s to $22.2$ s in the \textit{Sup-US} setting, from $42.3$ s to $29.9$ s in the \textit{Unsup-US} setting, and from $51.5$ s to $33.7$ s in the \textit{Sup-UI} setting.

\par In principle, the choice of pretrained decoder should have a smaller impact on the MSM since it only serves as an initialization, whereas the HMM post-processing directly operates on the decoder's output. However, the results do not show that the MSM is substantially more robust in the \textit{Unsup-US} and \textit{Sup-UI} settings, where the pretrained decoder is typically weaker. This suggests that the MSM may be susceptible to local optima during EM iterations, meaning the initialization remains a significant factor in final performance. This is consistent with the low SNR of EEG data and the high dimensionality of the MSM parameter space.

\section{Conclusion}

\par We proposed a MSM for sample-level sAAD that integrates the relationship between EEG and speech envelopes with the temporal dynamics of attention. The MSM achieves decoding accuracy comparable to recently proposed HMM post-processing methods while yielding significantly shorter switch detection latency. Furthermore, it provides per-sample attention decisions rather than using a traditional window-based approach, eliminating the need to pre-select a window length. While the current formulation relies on a linear regression framework, it can, in principle, be extended to non-linear architectures such as deep neural networks. However, the estimation of such non-linear models is inherently challenging given the low SNR of EEG data and may require more advanced inference techniques, such as variational inference or Markov Chain Monte Carlo methods. We leave these explorations for future work.

\bibliographystyle{IEEEtran}
\bibliography{IEEEabrv, ref}

\end{document}